\documentclass[12pt,preprint]{aastex}

\usepackage{natbib}
\usepackage{graphicx}

\begin{document}

\newcommand{\lya}{Lyman-$\alpha$}
\newcommand{\eqw}{\hbox{EW}}
\def\erg{\hbox{erg}}
\def\cm{\hbox{cm}}
\def\sec{\hbox{s}}
\def\f17{f_{17}}
\def\Mpc{\hbox{Mpc}}
\def\Gpc{\hbox{Gpc}}
\def\nm{\hbox{nm}}
\def\km{\hbox{km}}
\def\kms{\hbox{km s$^{-1}$}}
\def\yr{\hbox{yr}}
\def\Myr{\hbox{Myr}}
\def\Gyr{\hbox{Gyr}}
\def\deg{\hbox{deg}}
\def\arcsec{\hbox{arcsec}}
\def\microJy{\mu\hbox{Jy}}
\def\micron{ \ifmmode {\mu{\rm m}}\else
               $\mu$m\fi}
\def\zre{z_r}
\def\fesc{f_{\rm esc}}

\def\ergcm2s{\ifmmode {\rm\,erg\,cm^{-2}\,s^{-1}}\else
                ${\rm\,ergs\,cm^{-2}\,s^{-1}}$\fi}
\def\flamunit{\ifmmode {\rm\,erg\,cm^{-2}\,s^{-1}\,{\rm \AA}^{-1}}\else
                ${\rm\,ergs\,cm^{-2}\,s^{-1}\,{\rm \AA}^{-1}}$\fi}
\def\fnuunit{\ifmmode {\rm\,erg\,cm^{-2}\,s^{-1}\,{\rm Hz}^{-1}}\else
                ${\rm\,ergs\,cm^{-2}\,s^{-1}\,{\rm Hz}^{-1}}$\fi}

\def\ergsec{\ifmmode {\rm\,erg\,s^{-1}}\else
                ${\rm\,ergs\,s^{-1}}$\fi}
\def\kmsMpc{\ifmmode {\rm\,km\,s^{-1}\,Mpc^{-1}}\else
                ${\rm\,km\,s^{-1}\,Mpc^{-1}}$\fi}
\def\cMpc{\ifmmode {cMpc}\else
                ${cMpc}$\fi}
\def\kpc{{\rm kpc}}
\def\oii{[O{\sc II}] $\lambda$3727}
\def\oiipair{[O{\sc II}] $\lambda \lambda$3726,3729}
\def\oiii{[O{\sc III}] $\lambda$5007}
\def\oiiipair{[O{\sc III}]$\lambda \lambda$4959,5007}
\def\taulya{\tau_{Ly\alpha}}
\def\taubar{\bar{\tau}_{Ly\alpha}}
\def\llya{L_{Ly\alpha}}
\def\ldlya{{\cal L}_{Ly\alpha}}
\def\nbar{\bar{n}}
\def\Msun{M_\odot} 
\def\mdyn{M_{dyn}}
\def\vmax{v_{max}}
\def\sqamin{\Box'}
\def\l43{L_{43}}
\def\ls{{\cal L}_{sym}}
\def\snrat{\ifmmode {\cal S / N}\else
                   ${\cal S / N}$\fi}
\def\siglos{\sigma_{\hbox{los}}}
\def\asf{\alpha_{SF}}
\def\bsf{\beta_{SF}}
\def\SFR{\hbox{SFR}}
\def\rhoeff{\bar{\rho_e}}
\def\MsunYr{M_\odot\,\hbox{yr}^{-1}}
\def\highz{{PEARS-N-101687}}
\def\highzRADec{{12:37:25.65 +62:17:43.5 (J2000)}}
\def\Netsig{{\cal N}}

\title{
A Lyman Break Galaxy in the Epoch of Reionization from HST Grism Spectroscopy
}

\author{
James E. Rhoads\altaffilmark{1},
Sangeeta Malhotra\altaffilmark{1},
Daniel Stern\altaffilmark{2},
Mark Dickinson\altaffilmark{3},
Norbert Pirzkal\altaffilmark{4},
Hyron Spinrad\altaffilmark{5},
Naveen Reddy\altaffilmark{6}, 
Nimish Hathi\altaffilmark{7},
Norman Grogin\altaffilmark{4},
Anton Koekemoer\altaffilmark{4},
Michael A. Peth\altaffilmark{4,8}, 
Seth Cohen\altaffilmark{1},
Zhenya Zheng\altaffilmark{1},
Tamas Budavari\altaffilmark{8},
Ignacio Ferreras\altaffilmark{9}, 
Jonathan P. Gardner\altaffilmark{10},
Caryl Gronwall\altaffilmark{11}, 
Zoltan Haiman\altaffilmark{12},  
Gerhardt Meurer\altaffilmark{14}, 
Leonidas Moustakas\altaffilmark{2},
Nino Panagia\altaffilmark{4,15,16},
Anna Pasquali\altaffilmark{17}, 
Kailash Sahu\altaffilmark{4}, 
Sperello di Serego Alighieri\altaffilmark{18}, 
Amber Straughn\altaffilmark{10},
Rachel Somerville\altaffilmark{19},  
Jeremy Walsh\altaffilmark{20}, 
Rogier Windhorst\altaffilmark{1},
Chun Xu\altaffilmark{21}, 
Haojing Yan\altaffilmark{22}
}

\begin{abstract}
  We present observations of a luminous galaxy at $z=6.573$ --- the
  end of the reioinization epoch --- which has been spectroscopically
  confirmed twice. The first spectroscopic confirmation comes from
  slitless {\it HST}\/ ACS grism spectra from the PEARS survey
  (Probing Evolution And Reionization Spectroscopically), which show a
  dramatic continuum break in the spectrum at restframe 1216 \AA. The
  second confirmation is done with Keck + DEIMOS. The continuum is not
  clearly detected with ground-based spectra, but high wavelength
  resolution enables the \lya\ emission line profile to be
  determined. We compare the line profile to composite line profiles
  at z=4.5. The \lya\ line profile shows no signature of a damping
  wing attenuation, confirming that the intergalactic gas is ionized
  at z=6.57.  Spectra of Lyman breaks at yet higher redshifts will be
  possible using comparably deep observations with IR-sensitive
  grisms, even at redshifts where \lya\ is too attenuated by the
  neutral IGM to be detectable using traditional spectroscopy from the
  ground.
\end{abstract}

{\small
\noindent ~~\\
~~\\
 $^{1}$ ASU (James.Rhoads@asu.edu) ~~~
$^2$ JPL ~~~  $^3$ NOAO    ~~~ $^4$ STScI ~~~ $^5$ UC Berkeley ~~~~~~~ \\
$^6$ UC Riverside ~~~  $^7$ OCIW ~~~ $^8$ Johns Hopkins ~~~ $^9$ MSSL ~~~
$^{10}$ NASA GSFC ~~~ $^{11}$ Penn State ~~~ $^{12}$ Columbia
~~~ $^{14}$ ICRAR ~~~ $^{15}$ INAF - Catania ~~~ $^{16}$ Supernova Ltd.
~~~ $^{17}$ U. Heidelberg ~~~\\ $^{18}$ INAF - Arcetri 
~~~ $^{19}$ Rutgers ~~~ $^{20}$ ESO ~~~ $^{21}$ Shanghai Inst. of Technical
Physics ~~~ \\$^{22}$ U. Missouri
}

\keywords{
 galaxies: high-redshift --- galaxies: formation --- galaxies: evolution
}

\section{Introduction}
To properly understand the history of cosmic dawn, we must
be able to reliably identify galaxies observed during the epoch of
reionization.  Such galaxies are the most likely sources of
the radiation that ionized intergalactic hydrogen.  They are the best
places to look for signatures of primordial star formation: even
if the buildup of heavy elements is rapid, the fraction of
galaxies forming their first generations of stars should be higher
if we observe them when the universe itself was young. 
The pace of their growth depends on incompletely understood
physical processes --- both the onset of star-formation in 
low-metallicity conditions, and the potential disruption of
later star-formation by 
the ionizing radiation and/or supernovae produced by the first 
stellar generation.  The best way to constrain the range of possible outcomes
from these various processes is to take a direct, observational
census of galaxies throughout the reionization era --- from its end
at $6 \la z \la 7$, back to the earliest galaxies we can identify.

Much progress has been made recently in this direction, thanks
primarily to the dramatic increase in near-infrared imaging
sensitivity and survey efficiency afforded by the Wide Field Camera 3
(WFC3) Infrared (IR) channel on the {\it Hubble Space Telescope
  (HST)}.  Imaging surveys with WFC3-IR have provided tens to hundreds
of $z>7$ galaxy candidates, identified by the Lyman-$\alpha$
absorption break in their broad band colors
\citep[e.g.,][]{Bouwens10,Yan10,Yan12,Finkelstein12}.  (We will refer
to these as ``Lyman break galaxies,'' while noting that selection by a
strong continuum break can identify either the 912\AA\ break due to
Lyman continuum absorption, or the 1216\AA\ break due to \lya\
absorption.  Since the \lya\ forest is optically thick for $z\ga 5$,
surveys for $z>5$ galaxies use the \lya\ absorption break, while those
at $z\la 3$ primarily identify the 912\AA\ break.)  These broad band
{\it HST} searches have broken new ground, 
primarily because the NIR sky is orders of magnitude darker in space.
Alternative, ground-based search methods can find
\lya\ emitting galaxies efficiently at selected redshifts ($z =$ 6.5,
6.9, 7.3, 7.7, 8.8) where the line falls in dark windows in the night
sky spectrum,  using either narrow bandpass imaging
\citep[e.g.,][]{Hu02,Rhoads04,Iye06,Willis08,Hu10,Ouchi10,Hibon10,
Tilvi10,Kashikawa11,Clement12,Shibuya12,Rhoads12,Krug12},
or direct spectroscopic searches \citep[e.g.,][]{Kurk04,Martin04,
vanBreukelen05,Martin08,Dressler11}.

However, issues remain.  Ground-based
near-IR spectroscopy can only confirm these objects easily when they
have strong \lya\ lines in clean regions of the night sky spectrum.
Thus, while dozens have been confirmed up to $z=6.5$ \citep{Hu10,Ouchi10,Kashikawa11}, only a handful are confirmed at higher redshifts \citep{Iye06,Rhoads12,
Shibuya12,Pentericci11,Ono12,Schenker12}.
The crucial \lya\ line may be rare and/or weak at redshifts where the IGM
was mostly neutral (and hence able to scatter \lya\ photons).
Meantime, sample contamination by foreground galaxies becomes
an increasing worry at higher
redshifts, where the volume available for such contaminants
becomes large.  Finally, the candidate lists from the highest
redshift galaxy surveys can be disturbingly unstable, showing little
overlap when different groups examine the same data, or even when
the same group re-observes the same field \citep[e.g.][]{Yan12,Oesch12a}.  

Slitless spectroscopy with the {\it Hubble Space Telescope}\/ offers a
solution to many of these issues.  Space telescopes avoid the
crippling effects of Earth's atmosphere on the near-IR sky.
{\it HST}'s spatial resolution is well matched to the sizes of high
redshift galaxies.  These slitless grisms thus provide unmatched 
sensitivity to continuum emission from faint, compact high
redshift galaxies.

Here we present the highest redshift spectroscopic
confirmation to date from {\it HST}'s slitless grisms: The galaxy
\highz, which was identified based on its Lyman-$\alpha$ break at
$z\approx 6.6$ in deep ACS G800L slitless spectra from the PEARS
survey (``Probing Evolution and Reionization Spectroscopically'').
While the sensitivity of the ACS grism declines at 9600, the WFC3 IR
channel has a similar grism and can perform spectroscopic
confirmations at even higher redshifts.

We organize the paper as follows.  In section~\ref{sec:pears}, 
we describe the PEARS survey observations and data analysis.
In section~\ref{sec:keck}, we present followup spectroscopy at
higher spectral resolution from the Keck telescope. In section~\ref{sec:discuss},
we discuss the implications of our findings for the field of high-redshift
galaxy searches.  We conclude in section~\ref{sec:summary}. 
Most photometry discussed 
here is from the {\it HST}-GOODS survey \citep{Giavalisco04}.  We
denote the GOODS filters F450W as ``$B_{450}$,'' F606W as ``$V_{606}$'',
F775W as ``$i_{775}$'', and F850LP as ``$z_{850}$,'' and use the AB 
magnitude system.
Throughout the paper, we adopt a $\Lambda$-CDM ``concordance
cosmology''  with $\Omega_M = 0.27$, $\Omega_\Lambda = 0.73$, and
$H_0 = 71 \kmsMpc$ \citep[see][]{Spergel07}.

\section{PEARS Grism Observations}\label{sec:pears}
PEARS is the most extensive systematic survey conducted with
the G800L grism on the {\it HST's}\/ Advanced Camera
for Surveys-Wide Field Camera (ACS-WFC).  PEARS is an {\it HST}\/ Treasury
program led by S. Malhotra (program ID HST-GO-10530).  It covers a total of
nine fields, including one deep pointing in the {\it Hubble} Ultra Deep Field,
and eight wide-field pointings (four each in the GOODS-North and GOODS-South
regions).  Each pointing was observed at three or four distinct roll
angles, to mitigate the impact of overlap between spectra of nearby
objects.  

The {\it HST} slitless spectra were reduced using the aXe package\citep{Kuemmel09},
following closely the procedure used for the earlier
GRism ACS Program for Extragalactic Science (GRAPES) survey \citep{Pirzkal04}.
For each roll angle, the relative offsets of all exposures
were determined using zero-order images and narrow emission lines.
The data for each roll angle were ultimately combined into 2D spectroscopic
stacks and extracted 1D spectra for each source and each observed position
angle.  

To identify and spectroscopically confirm the highest redshift Lyman
break galaxies in the survey, we followed a procedure based on
\citet{Malhotra05}.  We started with the GOODS v1.9 images
and performed our own SExtractor photometry.  We then applied a ``liberal''
$i$-dropout criterion to generate a list of candidate Lyman break galaxies.
Since the GOODS data do not include observations redder than $z$-band
(and our candidate selection was done prior to the installation of WFC3), 
this ultimately amounts to using $i_{775}-z_{850} > 0.9$~mag.  
For each of these objects, we calculated the net signifance (``netsig'') 
parameter $\Netsig$
\citep{Pirzkal04} to determine which spectra might have sufficient
information for a redshift measurement.  (``Netsig'' is defined by
first sorting all pixels in a spectrum in descending order of 
signal-to-noise ratio; calculating the signal-to-noise ratio
$S_n$ obtained by combining flux from the brightest $n$ pixels, 
for all $n$ between 1 and the total number of pixels in the spectrum;
and finally taking $\Netsig = \max\{S_n\}$.)

After selecting candidates by $(i-z)$ color and ranking them by netsig,
several PEARS team members (including SM, JER, NP, SC, \& NG) examined 
the spectra by eye.  We did this because a straight $(i-z)$
color cut can select extremely red objects (EROs) as well as Lyman break
galaxies (LBGs), but the spectral signatures of the two are distinct.
(EROs show a smooth rise towards the red, while LBGs at these redshifts
show a step function at the redshifted wavelength of 
\lya\ forest absorption.)

\begin{figure}
\plotone{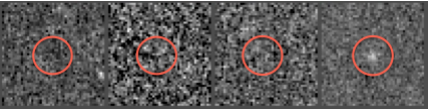}
\caption{Direct imaging of \highz\ from the GOODS survey
\citep{Giavalisco04}.  The object is effectively undetected in 
the $B_{450}$, $V_{606}$, and $i_{775}$ bands (left three panels), 
while it is clearly seen
in the $z_{850}$ band (right panel) with magnitude $z_{850} = 26.16$ (AB).  
Each panel is $1.5''\times 1.5''$ in size.
}
\label{fig:dirim}
\end{figure}

\begin{figure}
\hbox{\hspace{2.05cm}\includegraphics[width=12.3cm,keepaspectratio=true]{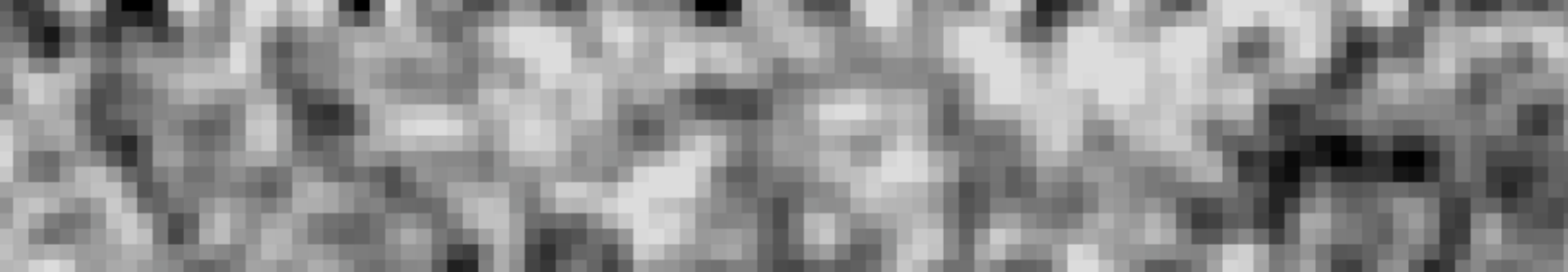}}
\includegraphics[width=15cm,height=4cm]{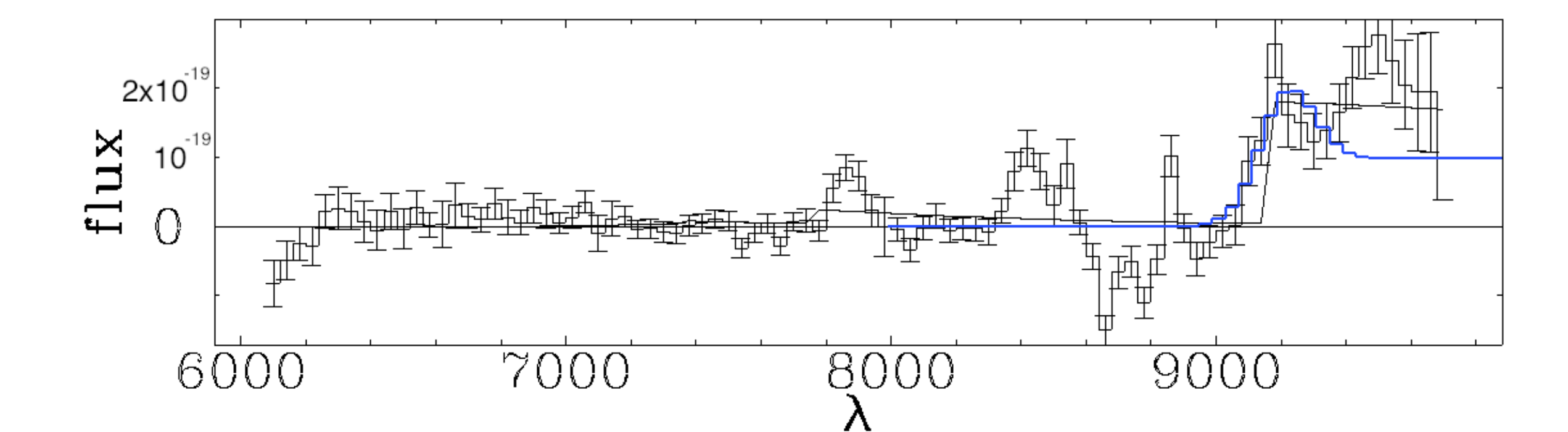}
\caption{{\it Upper:}\/ The 2D PEARS spectrum of \highz, displayed
in inverse video (dark = more flux),
with all three position angles coadded and with a 2 pixel ($0.1'' 
\times 80$\AA) FWHM 
smoothing applied.  A short segment of continuum is evident,
with a blue edge near 9230\AA\ due to the Lyman-$\alpha$ forest 
break, and a red cutoff imposed by the falloff of instrumental efficiency.
{\it Lower:} A 1D extraction of the PEARS spectrum, in 
\flamunit.  The thin black curve
is a best-fitting continuum + IGM absorption 
model with redshift $z=6.6$ and a continuum 
level of $AB=25.7$ mag on the red side of the \lya\ break. The blue curve
is a model based on the line flux from the Keck spectrum, and the continuum
flux level and line spread function width expected based on {\it HST} imaging.
}
\label{fig:grism}
\end{figure}

The highest redshift object identified in the PEARS-Wide fields
through this process was the galaxy \highz, at 
equatorial coordinates \highzRADec.  This object
has magnitude $z_{850} = 26.16$~mag,
while it is undetected in the $B_{450}$, $V_{606}$, and $i_{775}$ bands. 
Its grism redshift estimate is $z=6.6\pm 0.1$ based on the observed 
location of the Lyman $\alpha$ break.  We show postage stamp images
of the object from the GOODS data in Figure~\ref{fig:dirim},
and the 2D PEARS spectrum in Figure~\ref{fig:grism}.

\section{Keck Followup Observations}\label{sec:keck}
We selected several galaxies, including \highz, for followup
observations during a Keck observing run of three 
half nights on UT 2007 April 13--15, using the DEep Imaging
Multi-Object Spectrograph \citep[DEIMOS;][]{Faber03}.
These observations were part of a program of a deep field followup 
led by Spinrad, Stern, and Dickinson, using Keck telescope time 
from the University of California system.   
\highz\ was included on three slit masks during this run,
and also on one additional mask observed in early 2008. 
All observations used the 600 line grism. The observations
are summarized briefly in Table~\ref{tab:specobs}.

\begin{table}
\begin{tabular}{llllll}
Mask & Obs dates (UT) & N$_{\hbox{\scriptsize exp}} \times$ Duration & Total time & Conditions
& Comments\\
\hline
hdf07c & 2007 Apr 14 & 4 $\times$ 1800s & 7200 s & clear, $>1''$ seeing & Good\\
hdf07d & 2007 Apr 14--15  & 5 $\times$ 1800s & 9000 s & clear, $>1''$ seeing & Good \\
hdf07e & 2007 Apr 16 & 2820s & 2820s & clear, $1.4''$ seeing & Not useful \\
hdf08a & 2008 Mar 06 & 6 $\times$ $\sim$1800s & 10500 s &  & Good 
\end{tabular}
\caption{Log of DEIMOS observations of \highz.\label{tab:specobs}}
\end{table}

\begin{figure}
\plotone{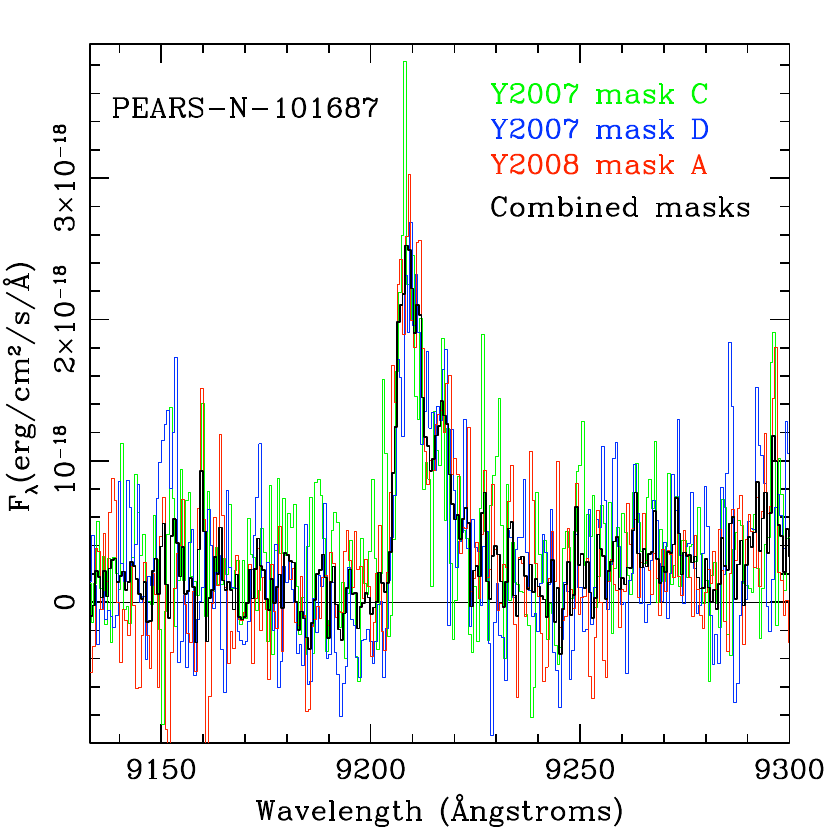}
\caption{The 1D Keck + DEIMOS spectrum of \highz.
The prominent line at 9210\AA\ is identified as \lya\ based on 
the strong break at 9210\AA\ in the {\it HST}\/ PEARS
grism spectrum, the complete absence of flux blueward of the line,
and the prominent asymmetry of the line.  The spectra from each individual
deep Keck+DEIMOS data set is shown with its own color, and the exposure-time
weighted mean of these three masks is shown in black. The spectra are
flux calibrated (with $\sim \pm 20\%$ accuracy) as described in the text.
There is a suggestion of continuum flux on the red side of the line
in the sum of all the Keck data.  No other convincing features are
seen in the full DEIMOS spectrum, which fully covers the 
range 5000--10000\AA.}
\label{fig:deimos}
\end{figure}

We show the extracted Keck + DEIMOS spectrum of \highz\ in
Figure~\ref{fig:deimos}.  The galaxy shows a prominent \lya\ 
line at redshift $z=6.573$, consistent with the grism-based redshift
of $z=6.6\pm 0.1$.  This line is plainly detected in each of the
three masks that have $> 1$ hour of exposure time.
Averaging the spectra from the three useful masks together
yields a better detection of the asymmetric line, along with hints
of the continuum on the red side of the line.
No other emission features are convincingly present in the
DEIMOS spectrum, which spans approximately 5000\AA\ $< \lambda < $10000\AA.

To estimate the spectroscopic line flux, we first calibrated the observed
spectra using a DEIMOS sensitivity function derived for the same grism but
on another night.  To compensate for differences in throughput (e.g., due to
differences in slit losses or atmospheric extinction between nights), we 
identified two objects in the 2008 mask with reasonably bright continuum flux
($i_{775} \sim 21$--$22$ mag).  For each, we weighted the 
spectrum by the throughput of the {\it HST} ACS F775W
filter, integrated, and compared the result to the broad band photometry
from the GOODS project.  This yielded a mean correction of about 30\%,
relative to our 
archival sensitivity function.
The reference objects used are both galaxies at moderate redshift, and 
although they are not point sources, they are considerably 
smaller than the DEIMOS slit.  We measured their fluxes in a virtual
$1''$ slit in ACS images both before and after smoothing with a $1''$
Gaussian ``seeing.''   We thereby found that differential slit
losses would require a 10\% correction to the spectroscopic flux
of a point source.  
Applying our calibration to the spectrum of \highz\ (and assuming the
source is pointlike in 1'' seeing), we find a spectroscopic line flux
of $f_{\rm{Ly\alpha}} \approx (2.8 \pm 0.6)\times 10^{-17} \ergcm2s$, where
the uncertainty is dominated by the flux calibration of the spectrum.

\section{Discussion}\label{sec:discuss}
\subsection{Other observations of \highz}
While no spectrum of \highz\ appears to have been previously
published, the object is listed as \lya\ candidate in \citet{Hu10} (Table
3, second entry), based on a narrowband excess in a filter with
9210\AA\ central wavelength and 120\AA\ FWHM.  
The narrow-band magnitude published in that work,
$AB=24.36$~mag, corresponds to a total flux of about $2.75\times 10^{-17}
\ergcm2s$ within the 120\AA\ filter.

The CANDELS survey \citep{Grogin11,Koekemoer11} provides WFC3-IR photometry 
of this region.  \highz\ is detected at high confidence in the
near infrared.  CANDELS catalog magnitudes for the source are
$z_{850} = 26.36 \pm 0.18$, $Y_{105} = 25.27 \pm 0.11$,
$J_{125} = 25.11 \pm 0.09$, and $H_{160} = 25.04 \pm 0.08$~mag. 
(Note the small [0.2 mag, $1\sigma$] difference between our 
previous $z_{850}$ photometry and the CANDELS project photometry.)
The $Y_{105}$ magnitude is near the wavelength of the \lya\
line, but unaffected by \lya\ forest absorption.
Using the $Y_{105}$ magnitude to estimate the continuum flux
density just redward of the \lya\ line,  we expect 
$\approx 6\times 10^{-18} \ergcm2s$ 
of continuum flux in the narrowband filter used by \citet{Hu10}.  This
leaves $\sim 2.1\times 10^{-17} \ergcm2s$ as the expected line flux 
based on photometry.  This is 25\% below the flux we derive from
the Keck DEIMOS observations, but consistent within the
combined uncertainties of all the data sets involved. 

\subsection{Comparison of grism and slit spectra}
The galaxy \highz\ is reminiscent of the bright $z=5.83$ dropout 
galaxy  UDF2225 \citep{Malhotra05} 
= SiD2 \citep{Dickinson04} = SBM3 
\citep{Stanway03,Stanway04}, in that the ACS grism spectrum shows a
clear continuum with a Lyman break, while the followup slit spectrum
from Keck shows a prominent \lya\ line.  This contrast is a
consequence of the differing capabilities of the two instruments.  The
high spatial resolution and low sky background of {\it HST}\/ + ACS provide
exquisite sensitivity to faint continuum emission.  On the other hand,
the higher spectral resolution of Keck+DEIMOS slit spectra provides a
clearer look at the \lya\ line.  The PEARS ACS spectrum does have
sufficient sensitivity to detect the observed \lya\ line, but the line
is blended with the Lyman break at the resolution of the grism.  For
the \lya\ line to appear obvious at $z \ga 5$, where the \lya\
forest is optically thick, its observer-frame equivalent width must
exceed the instrumental resolution, which is about 150--200\AA\ for
\highz\ (based on its half-light diameter in the GOODS images).
The observed equivalent width is modestly larger than this threshold, but
not so large that the line is expected to be prominent in the grism
spectrum.  The 2D HST spectrum also shows a hint of extended \lya\ emission
\citep[see][]{Rhoads09,Bond10,Finkelstein11b},
which would lie outside the extraction region for PEARS 1D spectra, 
but within the wider Keck slit, further increasing the relative prominence
of the line in the Keck spectrum.

We have modeled the expected 1D grism spectrum by assuming a flat
continuum at the level of the $Y_{105}$ flux measurement, a \lya\ line
at the wavelength and flux observed by Keck, and a line spread
function determined by the observed $z_{850}$ angular size of the
object and the dispersion of the ACS G800L grism.  The resulting model
is shown as a blue curve in figure~\ref{fig:grism}.  Near the \lya\
line, from 9000--9400\AA, the agreement is quite good.  The largest
discrepancy is at yet redder wavelengths, where the grism spectrum
continuum appears to be above the $Y_{105}$ flux.  This may be 
due to an edge effect always present in flux-calibrated slitless
spectra of extended sources.  The counts at a particular pixel include
redder light from one edge of the source, and bluer light from the
other edge, but all counts in the pixel are converted to flux density 
using a single system throughput, which is the one appropriate for light
from the centroid of the source that is dispersed onto that pixel. Where
the efficiency is changing rapidly with wavelength, as it does in the
9500\AA\ region for the ACS/WFC, the net effect is an
over-estimate of the red flux.
(Bumps in the 1D ACS spectrum at 7850\AA\ and 8400\AA\  
could be contamination by other fainter sources, transmissive gaps in the 
OH forest absorption, or simply regions of somewhat correlated noise
in the extracted spectrum.)

\subsection{\highz\ in context}
The galaxy \highz\ is, for its redshift, a moderately bright object.
It has a 1500\AA\ absolute magnitude $M_{1500} =
-21.38$~mag, based on an interpolation of the CANDELS $Y_{105}$ and
$J_{125}$ fluxes.  Compared to the published luminosity function for a
sample of candidate $z\approx 6.6$ LBGs from \citet{Bouwens11},
\highz\ is about $3\times$ (or 1.25 magnitudes) brighter than $L^*$
(the characteristic galaxy luminosity for the best fit Schechter
function).

The \lya\ luminosity of \highz\ is $\llya = 1.4\times 10^{43}
\ergsec$, based on its line flux from the DEIMOS spectrum and a
``concordance'' cosmology luminosity distance of $d_L = 65.5 \Gpc$.
This is on the bright end of the distribution for $z=6.5$
narrowband-selected \lya\ galaxy samples, which yield Schechter
function fits with characteristic luminosities of $L_* =4.4\times
10^{42}$, $5.8 \times 10^{42}$, and $1.0\times 10^{43} \ergsec$,
respectively, for \citet{Ouchi10}, \citet{Kashikawa11}, and
\citet{Hu10}.

The spectroscopic line flux of \highz, combined with its $Y_{105}$
flux density, yields an observer frame equivalent width of $EW = 290 \pm
80$\AA, or in the rest frame, $EW_{0} = 38 \pm 12$\AA.  This is below the
average for narrowband-selected samples, as one might expect given
that we identified the object by its continuum trace in the PEARS spectrum.
\lya\ galaxies often have rest frame equivalent widths above 200\AA\
\citep{Malhotra02}, and at $z=6.5$, over 75\% of the narrowband
selected \lya\ emitters have $EW_{0} > 40$\AA\ \citep{Kashikawa11,
Ouchi10}.

The \lya\ line asymmetry in \highz\ is prominent, even by the
standards of high redshift \lya\ emitting galaxies.  Using the asymmetry 
measures\footnote{These are $a_\lambda = (\lambda_{10,r} - \lambda_{p})
/ (\lambda_p - \lambda_{10,b})$ and $a_f = 
\int_{\lambda_p}^{\lambda_{10,r}} f_\lambda d\lambda / 
\int_{\lambda_{10,b}}^{\lambda_p} f_\lambda d\lambda$.  Here $\lambda_p$
is the wavelength where the line peaks, and $\lambda_{10,b}$ and 
$\lambda_{10,r}$ are the wavelengths where the flux falls to 10\% of peak 
on the blue and red sides of the line.}
from \citet{Rhoads03}, we find $a_\lambda = 3.59$ and $a_{f} = 3.47$.
Comparing these to a sample of 58 \lya\ emitters observed using
the same spectrograph and grating \citep{Dawson07}, 
\highz\ has the largest value of $a_f$, and the fifth-largest value
of $a_\lambda$.  This may be partly due to \lya\ forest absorption of flux
on the blue side of the systemic velocity, since the \lya\ forest optical depth
at $z=6.57$ will exceed that at the lower redshift ($z=4.5$) sample of 
\citet{Dawson07}.

In addition to its asymmetry, the line profile shows a dip at 9214\AA,
separating a secondary peak (at 9217\AA) from the primary one (at
9207\AA).  The dip is about 4\AA\ wide; the flux density there drops
by about 40\% relative to a smoothed ``envelope'' of the line flux;
and the feature is significant at about the $3\sigma$ level.  Such a
feature could be explained by neutral gas in front of the emitter; two
distinct emitting regions; or some more complex interplay of \lya\ emission 
and scattering in a moving medium.  If we interpret the dip as 
absorption, we can estimate the equivalent width and column density
of the absorber by interpolating across the dip (from 9211.5\AA\ 
to 9217\AA).  This yields an equivalent width of $-1.8$\AA\ (observer-frame)
or $-0.24$\AA\ (rest frame; here the ``$-$'' sign indicates absorption).
The maximum optical depth is $0.6$, and the system is $100\kms$ wide (FWHM
in optical depth).  The required column density is only $4\times 10^{13} 
\cm^{-2}$ of neutral gas--- comparable to a \lya\ forest absorber,
though with a greater velocity width.   The superimposition of this
absorption on the red part of the emission line would suggest an infalling
gas cloud in the neighborhood of \highz.

If we instead interpret the feature as additional emission at
$9217$\AA, the corresponding line flux is $2\times 10^{-18} \ergcm2s$.
This could correspond to a {\it faint} \lya galaxy along the
line-of-sight to \highz, with a redshift of $6.582$ (330 \kms\ redder
than \highz\ itself).  This would require a projected separation of
$\la 2 \kpc$ between \highz\ and the putative interloper, since there
is no visible evidence of a double morphology in the HST imaging.  The
line-of-sight separation could be much larger--- a reasonable fraction
of $\Delta v / H(z) \sim 400 \kpc$.  Thus, the ``neighboring galaxy''
hypothesis requires precise alignment along the line-of sight.  We
prefer instead either the absorber hypothesis, other radiative
transfer effects internal to \highz, or perhaps just a $\sim 3\sigma$
statistical fluke in the spectroscopic observations.

\subsection{Implications for Reionization} 
We have compared the spectrum of \highz\ to a comparison sample 
of \lya\ emitting galaxies at redshift $z=4.5$ \citep{Dawson07}.
We selected these comparison spectra because they were
obtained with the same instrument.
We first constructed a composite \lya\ line spectrum, 
by multiplicatively rescaling both the flux densities and wavelengths
of individual galaxies' spectra to a common peak, and then computing
both the mean and variance at each pixel.

\begin{figure}
\plotone{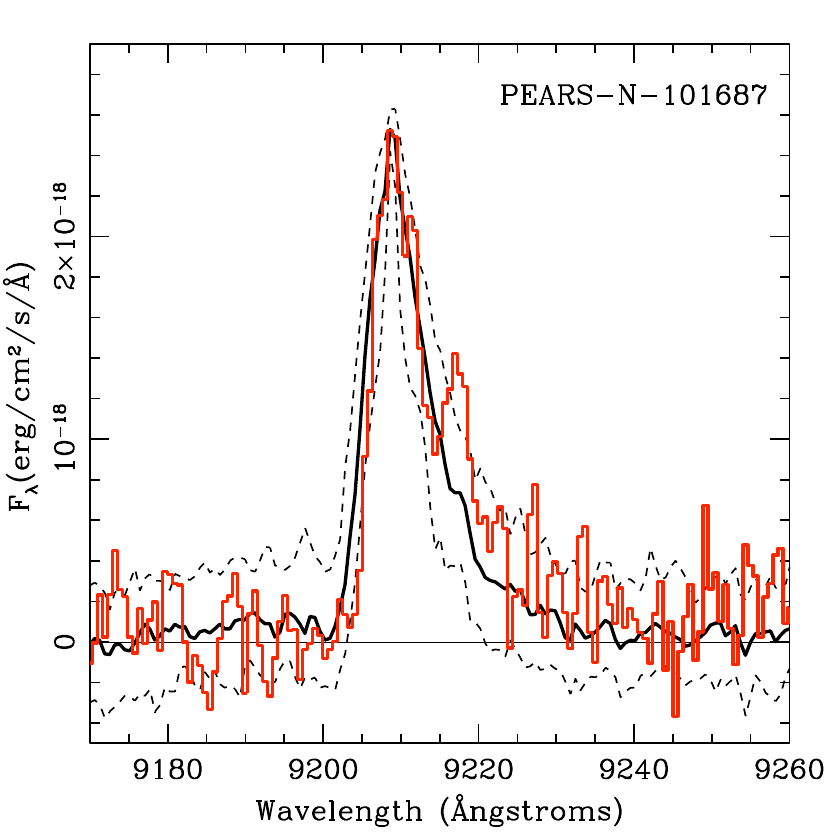}
\caption{The 1D Keck + DEIMOS spectrum of \highz\ (red histogram)
is plotted atop the composite spectrum of $\sim 70$ \lya\ galaxies
at redshift $z\approx 4.5$ \citep{Dawson07}.  The heavy black
line shows the composite spectrum, while the dotted black lines 
enclose the $\pm 1\sigma$ range for individual spectra.  The similarlity
of the \highz\ spectrum to the $z\approx 4.5$ composite argues
against major differences in the IGM ionization state between 
$z=4.5$ and $z=6.5$.  In particular, the 
sharp cutoff at the blue edge of the peak would be hard to reconcile
with the damping wing of neutral intergalactic gas.
}
\label{fig:composite}
\end{figure}

We plot both the composite $z=4.5$ \lya\ spectrum and the Keck
spectrum of in figure~\ref{fig:composite}.  The agreement between the
line profiles is remarkably good, apart from the dip/bump feature
discussed above.  Such agreement between $z\approx 6.57$ and $z=4.5$
would be destroyed by the damping wing of \lya\ if the intergalacic
medium around \highz\ were significantly neutral
\citep[see, e.g., fig.~5 of][]{Jensen13}.

This conclusion is also supported by  \lya\ luminosity function arguments.
Between 1 and 3 \lya\ emitting galaxies as bright as \highz\  might
be expected in the $z>6$ PEARS survey volume.  If we posit attenuation
of the \lya\ line by $50\%$, however, as expected for neutral
fractions $\ga 50\%$ \citep[e.g.][]{Haiman02,Jensen13}, 
the intrinsic line luminosity doubles.
Then, the expected number of galaxies plummets to $\sim 0.15$ in the
full PEARS survey volume.

\subsection{The need for spectroscopy}
Samples of well over 100 Lyman break selected galaxy candidates have
now been published for photometric redshifts $z>7$, based on 
the combination of 
deep photometry at optical and near-IR wavelengths using {\it HST}\/ 
\citep[e.g.,][]{Bouwens10,Yan10,Yan12,Finkelstein12}.
This represents substantial progress in understanding galaxy 
evolution around the end of the reionization era. 
{\it However, essentially 
{\bf all}\/ these objects remain candidates at the moment,
unconfirmed by spectroscopy.}   Deep grism spectra
from {\it HST}\/ can play a unique role in fixing this.

The overlap between samples published by different groups can be
distressingly small, even when those groups use exactly the same data
sets.  For example, consider recent publications on the bright end of the
$z\approx 8$ LBG luminosity function by \citet{Yan12} and
\citet{Oesch12a}, both using the first epoch of the CANDELS deep
observations of the GOODS-S region.  The two groups publish 8 and 9
candidate galaxies, respectively.  However, only two objects are identified
by both papers.  Similar levels of inconsistency have been frequent
in earlier studies. 
Indeed, a recent paper by \citet{Ellis13} reports that
{\it no} previously published galaxy candidate at $8.5<z<10$ 
in the Hubble Ultra Deep Field remains a viable high-redshift
object after the addition of deeper imaging in the WFC3-IR 
F105W (``$Y_{105}$'') and F140W (``$JH_{140}$'') filters, 
while reporting a set of seven new candidates in that redshift range.

Several factors contribute to unreliable candidate lists.
First, candidates are generally sought down to the limit of the survey
depth, and objects near the faint limit of the data will
then inevitably outnumber brighter, better measured sources.  
This means that photometric noise can push a galaxy across the
selection line --- in {\it or} out of a candidate sample, either in
brightness or in color.  Apparently minor differences in the choice
of photometry method (apertures of various radii vs. SExtractor ``magauto'';
different methods of sky background estimation; etc) can thus change samples
appreciably.  

Additionally, different authors may choose somewhat different criteria 
in selecting their candidates.  Some use photometric redshifts
\citep{Finkelstein12}, while most others use straight color and magnitude
cuts \citep[e.g.,][]{Bouwens10,Oesch12a,Yan12}, 
but the adopted color cuts are not always the same.  

As the search redshift increases, the volume of foreground space and
the variety of possible foreground contaminants increases too.  For
galaxies at $z>7$, plausible foreground contaminants include Galactic
brown dwarfs, and both early-type galaxies and ultra-strong emission
line sources at intermediate redshifts 
\citep[e.g. the candidate lensed $z=11$ galaxy 
A2667-J1, whose spectrum revealed it to be an \oiiipair\ emitter 
at $z=2.082$;][]{Hayes12}.  
While published $z>7$ candidates
usually have spectral energy distributions that are less well fit by
any foreground model than by a Lyman break galaxy at $z>7$, a majority
allow viable $z \ll 7$ solutions. Recently, 
\citet{Pirzkal13} have carried out an analysis of redshift estimation based on Markov Chain Monte Carlo fitting (MCMC) for samples of $ z > 8$ galaxy candidates from the HUDF. They estimate that there is an average probability of 21\% that these sources are low redshift interlopers.

Spectroscopic followup of $z>7$ candidates can resolve these
uncertainties.   Ground-based spectra can confirm true $z>7$ galaxies, but
generally only when they have strong \lya\ emission that is neither blocked
by atmospheric H$_2$O absorption nor blended with strong OH airglow
lines at the resolution of the spectrograph.  Likewise, ground-based
spectra may definitively rule out foreground objects whose Lyman-break
colors are due to strong emission lines in one or two filters.

However, sources without strong emission lines require continuum
spectroscopy for definitive confirmation.  For redshifts $z<6$,
a minority of LBGs have strong \lya\ emission
\citep{Steidel00,Stark10}.  As we push to higher redshifts, within
the epoch of reionzation,  \lya\ will be obscured by 
resonant scattering in an increasingly neutral intergalactic medium
(IGM) \citep{Miralda98,Haiman99,Rhoads01}.  This effect offers valuable
tests of reionization 
\citep[e.g.,][]{Malhotra04,Stern05,Pentericci11,Ono12,Schenker12}. 
It also means that {\it continuum break spectroscopy is crucial} 
for spectroscopic confirmations in the epoch of neutral gas.

Such spectroscopy is impractical from the ground with current instruments.
From space, the absence of OH emission lines and water absorption makes
the job much easier.  To obtain a Lyman break confirmation at the 
knee of the luminosity function ($L^*$), we need to detect the continuum
with good statistical significance, over a wavelength range of several hundred 
\AA{}ngstrom on the red side of the Lyman break.  
Another way to test the IGM neutral fraction is to look for evolution
in the fraction of LBGs that show \lya\ line emission
\citep[e.g.][]{Stark10}.  Multiple groups have studied this with small
sample sizes, and find some evidence for a decreasing fraction at z
$\ge$ 7.0 \citep{Pentericci11,Schenker12,Caruana12} though the
conclusions of \citet{Ono12} are more ambiguous.  A significant
concern here is that {\it only} the \lya\ emitting galaxies have so
far been confirmed spectroscopically.  The candidate galaxies not
confirmed with a \lya\ line will include those that lack line
emission, but could also include cases where a \lya\ line is hidden
behind atmospheric H$_2$O absorption or OH emission features, or
interlopers that are not actually high-redshift sources at all.  Thus
when the \lya\ fraction is reported the denominator of high-z galaxies
is itself in doubt.  The papers discussed above make statistical
corrections to account for these effects, but such corrections require
an exquisite understanding of photometric redshift uncertainties.
Space grism spectroscopy can provide direct spectroscopic confirmation
based on continuum break.  Deep integrations with {\it HST}\/ grisms
can accomplish this, spectroscopically confirming redshifts for
objects as faint as 27th magnitude \citep[this
work;][]{Malhotra05,Rhoads09}.

\section{Summary/Conclusions}\label{sec:summary}
We present here \highz, the highest redshift galaxy identified in the
wide-field component of the PEARS slitless spectroscopic survey, which
achieved a depth of 20 orbits of {\it HST}\/ ACS G800L slitless spectroscopy
over 80 square arcminutes.  This galaxy, with a grism redshift of 
$z=6.6 \pm 0.1$,
remains the highest redshift object yet confirmed with {\it HST}\/ slitless
spectroscopy.  While fainter objects have been spectroscopically confirmed
at the sensitivity limit of our ACS slitless spectroscopy in the
Hubble Ultra Deep Field, the wavelength limit of the ACS Wide Field
Camera CCD detectors effectively limits the survey redshifts to $z \la
6.7$ \citep[][]{Malhotra05}.

Keck telescope + DEIMOS followup of this object confirms and refines
its grism redshift: $z=6.573$.  The object has a rest-frame UV continuum
magnitude $M_{1500} = -21.38$~mag, a \lya\  line luminosity of
$1.4 \times 10^{43} \ergsec$, and a rest-frame equivalent
width of $38 \pm 12$\AA.  This makes it a relatively luminous \lya\ emitting
Lyman break galaxy.  We
emphasize that the discovery and redshift from {\it HST}\/ PEARS are based
primarily on the continuum and \lya\ forest break, and not on the
\lya\ emission line, which is not prominent in the {\it HST}\/ spectrum.
The Keck spectrum matches closely the composite spectrum of $z\approx 4.5$ 
\lya\ emitting galaxies.  In particular, it shows no evidence for additional
absorption at the blue edge of the line, as would be expected
in a significantly neutral IGM.  We therefore conclude that reionization
is essentially complete at $z\approx 6.6$ in the neighborhood of \highz.

The discovery of this object demonstrates the value of deep continuum
observations with {\it HST}\/ slitless grisms for spectroscopic confirmation
of galaxies in the epoch of reionization.  Comparably sensitive
observations with the {\it HST}\/ WFC3-IR channel grisms have the potential to
provide Lyman break spectroscopic confirmations of $z > 7$ galaxies ---
something that still eludes our other observational capabilities, and
that now presents a large obstacle in advancing our understanding 
of galaxy evolution in the era of cosmic dawn.

\section*{Acknowledgments}
JER and SM thank the DARK Cosmology Centre and Nordea-fonden
in Copenhagen, Denmark, and Anri's Place in Betalbatim, India,
for their hospitality during the completion of this work.
We thank Mauro Giavalisco and the GOODS team for providing early
access to GOODS v1.9 and v2.0 images to help with spectroscopic
extractions.  We thank Emanuele Daddi for his
contributions to the PEARS project.
This work has been supported by grant HST-GO-10530 from STScI,
which is operated by AURA for NASA under contract NAS 5-26555.
The work of DS was carried out at Jet Propulsion Laboratory,
California Institute of Technology, under a contract with NASA.
The Institute for Gravitation and the Cosmos is supported by the
Eberly College of Science and the Office of the Senior Vice President
for Research at the Pennsylvania State University.
Some data presented herein were
obtained at the W. M. Keck Observatory.  The Observatory was made
possible by the generous financial support of the W. M. Keck
Foundation.
The authors wish to recognize and acknowledge the very significant
cultural role and reverence that the summit of Mauna Kea has always
had within the indigenous Hawaiian community. We are most fortunate to
have the opportunity to conduct observations from this mountain.
%




\end{document}